# Real spin glasses relax slowly in the shade of hierarchical trees

E. Vincent[1], J. Hammann, M. Ocio

*Service de Physique de l'Etat Condensé (CNRS URA 2464), IRAMIS/SPEC, CEA Saclay, 91191 Gif sur Yvette Cedex, France*

Abstract

The Parisi solution of the mean-field spin glass has been widely accepted and celebrated. Its marginal stability in 3d and its complexity however raised the question of its relevance to real spin glasses. This paper gives a short overview of the important experimental results which could be understood within the mean-field solution. The existence of a true phase transition and the particular behaviour of the susceptibility below the freezing temperature, predicted by the theory, are clearly confirmed by the experimental results. The behaviour of the complex order parameter and of the Fluctuation Dissipation ratio are in good agreement with results of spontaneous noise measurements. The very particular ultrametric symmetry, the key feature of the theory, provided us with a simple description of the rejuvenation and memory effects observed in experiment. Finally, going a step beyond mean-field, the paper shortly discusses new analyses in terms of correlated domains characterized by their length scales, as well as new experiments on superspin glasses which compare well with recent theoretical simulations.

Keywords: spin glass, relaxation, noise, aging, rejuvenation, memory, mean field, hierarchical tree

[1]Corresponding author, eric.vincent@cea.fr

# 1. Introduction

The theorist's spin glass is a set of randomly interacting objects, a problem whose conceptual simplicity sharply contrasts with the sophistication of the mathematical methods that have been developed to solve its mean-field approximation [sherrington75,mezard87]. The experimentalist's spin glass is usually obtained by randomly diluting magnetic ions in a 3-dimensional compound. The historical example is that of intermetallic alloys, like for instance Cu:Mn$_{3\%}$, in which 3% of (magnetic) Mn atoms are thrown by random in a (non-magnetic) Cu matrix. The Mn magnetic atoms sit at random positions, therefore are separated by random distances, and the oscillating character of the RKKY interaction with respect to distance makes their coupling energy take a random sign.

Later on, spin glasses have been identified within insulating compounds. An example is the Indium diluted Chromium thiospinel $CdCr_{2x}In_{2(1-x)}S_4$ [alba82], with superexchange magnetic interactions between the Cr ions. The nearest neighbour interactions are ferromagnetic and dominant for x=1 (a ferromagnet with $T_c$=80K), but the next-nearest ones are antiferromagnetic. Hence, when some (magnetic) Cr ions are randomly substituted by (non-magnetic) In ions, some chromium sites experience a net antiferromagnetic coupling. The balance that globally favours ferromagnetism for zero or small In-dilution is disturbed, and the ferromagnetic phase is replaced by a spin-glass phase for x≤0.85.

Interestingly, many generic features have been found in the very different chemical realizations of spin glasses that have been studied. Metallic as well as insulating 3d spin-glasses show a well defined phase transition at $T_g$ (attested by the critical behaviour of some quantities), and slow dynamics is observed in the spin-glass phase with the occurrence of such amazing phenomena as aging, rejuvenation and memory effects. Some systematic differences have indeed been found as a function of spin anisotropy, but they are not obviously related to the metal/insulator character or to any other chemical feature [bert04].

Many of the observed features of experimental spin glasses correspond well with some aspects of the famous Parisi solution [mezard87] of the mean-field spin glass problem [sherrington75]. In spite of its complexity and of the oversimplifications made in its interpretation, the Parisi solution with the "replica symmetry breaking" (RSB) and the resulting ultrametricity indeed provides a very general qualitative background for the unusual properties of the aging effects of real spin glasses.

We discuss here a few of the specific experimental results which could be nicely accounted for in the light of the Parisi solution, affording an important step forward in the overall clarification of the spin glass problem.

# 2. A few experimental facts at the light of mean-field spin-glass results

## 2.a FC and ZFC susceptibilities

For the mean-field spin glass, two susceptibilities can be defined: (i) $\chi_{LR}$ which corresponds to the linear response of the system trapped below $T_g$ in a specific pure state, and (ii) $\chi_{eq}$ which is the equilibrium susceptibility thermally averaged over the whole phase space. The difference of the two susceptibilities is the hallmark of replica symmetry breaking [mezard87,parisi06]. This difference is illustrated in Fig. 1a. It compares well with the experimental results shown in Fig.1b, where $\chi_{zfc}$ (the zero field cooled susceptibility) represents the response to a small field applied on a sample that has been previously cooled in zero field, and $\chi_{fc}$ (the field cooled susceptibility) stands for the susceptibility observed while cooling the sample from above $T_g$ in a constant small field. $\chi_{zfc}$ behaves like the theoretical $\chi_{LR}$ whereas $\chi_{fc}$ remains (almost) constant below $T_g$ as does $\chi_{eq}$. The similarity between experiments and analytic results in the mean-field approximation is remarkable.

It is however important to note that there are many other experimental situations in which irreversibilities appear below some freezing temperature, with a separation of FC and ZFC susceptibilities. In superparamagnets, for instance, the flipping of the superspins born by magnetic nanoparticles is impeded when the thermal fluctuations become too small compared to the anisotropy energy of the nanoparticles. This blocking mechanism is intrinsic to the individual nanoparticles, and freezing occurs even in the absence of any significant interactions between them. Due to size distribution effects, freezing occurs over a rather wide range of temperatures. Therefore the FC curve after its splitting from the ZFC one does not remain constant with decreasing temperature but continues to grow up, although much more slowly than the Curie-Weiss law observed above the freezing point. In contrast, in spin glasses, the FC susceptibility is *essentially flat* below the well-defined temperature $T_g$ at which it separates from the ZFC one (Fig.1b).

The existence of true phase transition has been proven by measurements of critical behaviours: divergence of the non-linear susceptibility [miyako79] and divergence of the characteristic response time in ac experiments [bontemps84] (see also [vincent86]). One easier experiment to check the cooperative process is the frequency dependence of the freezing temperature [tholence84]. In superparamagnets with non-interacting nanoparticles, this frequency dependence obeys a simple Arrhenius law, and the freezing temperature tends to *T=0* at infinite times. In spin-glasses, the frequency dependence is either fitted to a Vogel-Fulcher law or to a power law of the reduced temperature, in which cases the freezing temperatures extrapolate to a non-zero value at infinite times [tholence84].

Again, the existence of a true phase transition in 3d and the divergence of the non-linear susceptibility can be predicted from the Parisi solution, even though in that case the solution becomes marginally stable and the presence of a line of singularities (extending over the entire interval from $T_g$ down to 0) is suggested [mezard84,mezard87].

## 2.b Spontaneous magnetic fluctuations and Fluctuation-Dissipation ratio

Spin glasses never reach equilibrium, and dynamical functions like the time autocorrelation of magnetization (magnetic noise) or the response function (relaxation after a field change) cannot be reduced to one-time quantities. This is the *aging* phenomenon: for example, the time variation of the magnetization after a field change depends on the time spent before the field change (see details and references in [vincent07]). In equilibrated systems, the Fluctuation-Dissipation ratio (FDR) between autocorrelation and response is proportional to temperature through the general Fluctuation-Dissipation Theorem. In out-of-equilibrium glassy systems like spin glasses in the aging regime, the FDR does not simply correspond to the temperature of the sample. But, in some generic models, it has been proposed that the quantity defined by the FDR behaves like an "effective temperature", different from the temperature measured by a more usual thermometer [cugliandolo94,franz94]. Some experiments have been set up to measure this effective temperature, among which the spin glass experiments by D. Hérisson and M. Ocio [herisson02].

Magnetic noise measurements are a very difficult challenge. Firstly, the spontaneous magnetic fluctuations (noise measured with no applied field) are very weak. Their typical amplitude corresponds to the response obtained for a magnetic field of $10^{-7}$ Oe, while usual susceptibility measurements are made with a field of the order of 1 Oe. Also, since the dynamics depends on the time spent after cooling the sample (aging), averaging the results to increase the statistical accuracy implies repeating the whole run (starting from above $T_g$). Secondly, the measurements must focus on the low frequency range, in which strong out-of-equilibrium effects show up. But this low-frequency range is easily polluted by many external factors (all kinds of slow drifts, like day-night cycle). D. Hérisson and M. Ocio succeeded in performing such difficult experiments, and could characterize the Fluctuation-Dissipation ratio in various time regimes. In the equilibrium time regime, they could check that the FDR was in agreement with the temperature of the sample as measured by a thermometer. And, for the first time in a glassy system, the crossover between equilibrium and non-equilibrium time regimes could be clearly identified, yielding an *effective temperature* (defined from the FDR value) of *1.5 $T_g$* (experiment at *0.6 $T_g$*) to *4 $T_g$* (experiment at *0.9 $T_g$*). The final results are shown in Fig.2, and all details can be found in [herisson02].

The FDR has another very stimulating theoretical interpretation. This measurable quantity is indeed reflecting, in the mean-field approximation, the shape of the order parameter distribution P(q) (see [franz98]). This is pictured in Fig.3 (from [parisi06]) for 3 theoretical cases:
A) domain growth-like system, *$1/T_{eff}$ = 0*, horizontal line in Fig.3,
B) 1-step RSB type models, straight lines of slope *$1/T_{eff}$*,
C) continuous RSB models, smooth curve extrapolating from a *1/T* slope (equilibrium dynamics) to a horizontal slope in the limit of zero autocorrelation, case of the mean-field spin glass.
The results of Hérisson and Ocio [herisson02] clearly show that the spin glass does not behave like a disguised ferromagnet (case A), but presents a degree of complexity which pertains to "replica symmetry breaking" glasses (case B or C), even if there is still further work to be done on this subject for more thorough conclusions.

## 2.3 Rejuvenation and memory: a hierarchical tree in the experiments

Another indication of aging can be seen in the slow relaxation of both components of the ac susceptibility at a given (low) frequency. In the experiment presented in Fig.4a, the out-of phase component χ" of the ac susceptibility at 0.1Hz shows an example of "rejuvenation and memory" phenomena. Starting from above $T_g$ where χ"=0, the sample is cooled by steps of 2K, with a waiting time of ½ hour at each temperature. During each waiting time, χ" relaxes down (aging). And each time the temperature is lowered by one further step, a jump of χ" is observed, followed by another (renewed?) relaxation, a behaviour termed "rejuvenation". After reaching the lowest temperature, the sample is re-heated continuously while recording χ"; amazingly, the "memory" of each of the aging stages performed during cooling is revealed in shape of "memory dips".

In Fig.4b we present the hierarchical picture proposed by the Saclay group to account for these phenomena (and many other experimental results, see e.g. [sasaki02]) [refregier87]. The shown scheme had been first suggested by [dotsenko85] in a theoretical approach to the Parisi solution. This very simple picture sketches the effect of temperature variations in terms of a modification of the free-energy landscape of the metastable states (and not only of a change in the transition rates between them). At fixed temperature *T*, aging corresponds to the slow exploration by the spin glass of the numerous metastable states. When the temperature is decreased from *T* to *T −ΔT*, the free-energy valleys subdivide into smaller ones, separated by new barriers. Rejuvenation arises from the transitions that are now needed to equilibrate the population rates of the new sub-valleys. For large enough *ΔT* (and on the limited experimental time scale), the transitions can only take place between the sub-valleys, in such a way that the population rates of the main valleys are untouched, keeping the memory of previous ageing at *T*. Hence the memory can be retrieved when re-heating and going back to the *T*-landscape.

In the mean-field model of the spin-glass, it has been shown that rejuvenation and memory effects can also be expected in the dynamics [cugliandolo99]. But the nature of a possible link between this hierarchical tree of the *metastable* states as a function of *temperature* and the mean-field hierarchical tree of *pure* states as a function of their *overlap* is not clear. Having this question in mind, the group of R. Orbach and the Saclay group proposed experiments in which the hierarchical tree of the metastable states could be explored quantitatively [hammann92]. The principle is rather simple: compare 2 experiments with different thermal histories which yield the *same* relaxation curve. In experiment #1, starting from above $T_g$, the sample is cooled in a small field and kept at temperature $T-dT$ during a waiting time $t_w$, then the temperature is set to T, the field is cut, and the relaxation of the "thermo-remanent magnetization" (TRM) is recorded. In experiment #2 almost the same procedure is used, but the sample is kept at temperature T during a waiting time $t_w^{eff}$, which is chosen in such a way that one obtains exactly the same relaxation curve as in case 1 (for details of the exact procedure see [hammann92,bert04] or references in [vincent07]).

During the waiting times the system probes a certain sub-space of the total phase space bound by a maximum typical barrier B. If the observed relaxation curves measured at T in case 1 and in case 2 are exactly identical, it is supposed that the same sub-space has been probed at T during tweff and at T-dT during Tw. The typical barrier bounding this sub-space has a value $B(T)=k_B T.Ln(t_w^{eff}/\tau_0)$ at T and $B(T-dT)=k_B(T-dT).Ln(t_w/\tau_0)$ at T-dt. It is thus possible to derive an estimate of the barrier height variation between T and T-dT [hammann92].

A large series of such data has been taken with the Ag:Mn2.6% spin-glass sample. The results are shown in Fig.5 in terms of the thermal barrier variation *dB/dT*. We see that the thermal barrier variation is negative: the barriers increase when going to lower temperatures. Also, plotted as a function of the barrier height B, we observe that the data taken at different *T*'s define a unique curve, meaning that *dB/dT* is a function of *B* only (the temperature T remains an implicit variable of the plot, at different T's one has access to a different range of barriers heights).

The interesting information lies in the steep slope of this *dB/dT* vs *B* curve, which may be fitted for instance to a power law function : *dB/dT = a B$^6$*. Integration yields *B~ (T-T\*)$^{-1/5}$*, where *T\** is a free integration constant, the temperature at which the particular barrier under consideration diverges. The existence of aging effects at all temperatures below $T_g$ implies the existence of a continuous distribution of barrier heights at *any* temperature. Hence, the shape of *dB/dT* vs *B* strongly suggests that barriers diverge at any temperature below $T_g$ : as the temperature is lowered to $T_g$, the first divergence of a barrier takes place, separating phase space into mutually inaccessible regions (pure states). As the temperature continues to lower, more barriers diverge, and the number of pure states, developing from the hierarchically organized metastable states, continues to grow [hammann92].

This view of a spin glass phase which is critical at all temperatures below $T_g$ is consistent with the prediction of the Parisi solution of the presence of a continuous line of singularities below Tg and a theoretically suggested interpretation of the existence of a continuous sequence of micro-phase transitions [mezard84, mezard87].

Of course, data on free-energy barriers are limited to the narrow 25-35 $k_B T$ range by the ratio between the experimentally available time scales and the microscopic time scales. The above interpretation [hammann92] rests on a very strong assumption : namely on the possibility to extrapolate the results over a very extended range of times compared to the measured time scales. Furthermore it is very difficult to get a clear view of the nature of these diverging barriers in such a complicated phase space. As will be seen in the following, the analysis of more recent series of experiments led us to a presently preferred interpretation in terms of a single phase transition at $T_g$ and the complex growth of correlated regions defined by their characteristic length scale [bouchaud01,bert04]. In this interpretation the fast growing barriers correspond to the activation energy needed to flip these correlated regions.

# 3. Length scales and temperature microscope

## 3.a Spin glass

Beyond a description of aging, rejuvenation and memory effects in terms of metastable states, it is of course very intriguing to imagine what kind of spin arrangements allow such complex phenomena when the temperature is varied [tanaka05]. It is very natural, as proposed in the "droplet model" [fisher88], to consider that the spin-glass, initially in a random configuration after the quench, slowly builds up from neighbour to neighbour a spin-glass local order over larger and larger length scales.

Thinking of the multiple memory experiment in Fig.4, the observed rejuvenation and memory effects have some implications in terms of these dynamic length scales:

(i) Rejuvenation when going from $T$ to $T-\Delta T$ indicates that the spin-spin correlations growing at $T-\Delta T$ are different from those established at $T$ (for a discussion about a possible interpretation in terms of "temperature-chaos" [fisher88, yoshino], see [bouchaud01]).

(ii) If correlations extend up to a coherence length $L^*(T)$ during ageing at $T$, in order to preserve this memory of $L^*(T)$, ageing at $T-\Delta T$ should occur without changing significantly the correlations established at the scale $L^*(T)$, that is, $L^*(T-\Delta T) < L^*(T)$.

In practice, the independence of aging at length scales $L^*(T-\Delta T)$ and $L^*(T)$ is realized by a strong separation of the related *time* scales $\tau$ : $\tau(L,T-\Delta T) \gg \tau(L,T)$. This necessary separation of the aging length scales with temperature is the "temperature microscope" effect proposed by J.-P. Bouchaud [bouchaud01]: in an experiment like shown in Fig. 4, at each stage ageing should take place at different length scales $L^*_n < \ldots < L^*_2 < L^*_1$. This hierarchy of embedded length scales as a function of temperature is a "real space" equivalent of the hierarchy of metastable states in the "phase space".

An example of a system which presents such a hierarchy of length scales is that of an elastic line in presence of pinning disorder [balents96]. Here, frustration arises from the competition between elastic energy, which tends to make the line straight, and pinning energy, which tends to twist the line to match the pinning sites. As sketched in Fig.6 [bouchaud00], starting from a random configuration after a quench, after some aging at $T$ the line can be pictured as a fuzzy ribbon which is equilibrated over a length scale $L^*(T)$. At smaller length scales, the line continues to fluctuate between configurations which are roughly equivalent at temperature $T$ (thus seen as a fuzzy ribbon). However, when going from $T$ to $T-\Delta T$, the difference between the equilibrium populations of some of these configurations may become significant, and a new equilibration at shorter length scales $L^*(T-\Delta T) < L^*(T)$ must take place. These dissipative processes cause a rejuvenation signal. Meanwhile, processes at length scale $L^*(T)$ are frozen at $T-\Delta T$, and the memory of previous aging remains intact despite the rejuvenation processes, which occur at smaller length scales. In the case of a directed polymer in a random medium, studied in [sales02], evidence for a "temperature-chaos" has been found. In the case of spin glasses, simulations indicate that rejuvenation can be obtained in the absence of such a temperature-chaos [berthier02, berthier03].

This scheme is a good guideline for the mechanism of aging, rejuvenation and memory in spin glasses. But it is not clear to define what objects could play the role of pinned elastic lines. Experiments on disordered ferromagnets show that spin-glass dynamics is also observed in this case, most probably due to the pinning/depinning dynamics of the walls [dupuis02]. Slow dynamics in spin-glasses might well be explained by wall-like dynamics, but for now we cannot identify what are these walls, and what is the nature of the domains which are separated by these walls (interesting similarities can be found with the "sponge-like" excitations observed in simulations [krzakala00,lamarcq02]).

Experiments have been designed in order to have access to the dynamical coherence length that grows during aging in a spin glass [joh99,bert04]. The idea starts from the observation that the magnetization relaxation following a field change (as well in TRM as in ZFC procedure) becomes faster when a higher field amplitude is used. The inflection point of the relaxation curve, usually observed around $t_w$ in low fields, is found to shift towards shorter values $t_w^{eff}(H)$ when the field H increases. This acceleration of the relaxation is ascribed to the effect of the Zeeman energy which couples the field to the magnetization $M(N_S)$ of a group of $N_S(t_w)$ spins which are able to flip coherently after a time $t_w$ : $E_Z = M(N_S).H$.

Thinking again in terms of free-energy barriers associated to macroscopic response times, we may define $B = k_B T.Ln(t_w/\tau_0)$, the typical barrier which corresponds to the maximum relaxation rate (inflection point) at $t_w$ in a low field experiment. For an experiment with a stronger field variation, this barrier is decreased by the Zeeman energy, namely $B-E_Z = k_B.Ln(t_w^{eff}/\tau_0)$. Measuring $t_w^{eff}$ for various sets of $t_w$ and $H$, one can determine $E_Z = M(N_S).H$, and with some assumption on the $N_S$ dependence of M [joh99,bert04] we can deduce $N_S(t_w)$, the number of correlated spins after time $t_w$ after the quench.

Fig.7 shows the $N_S$ results from a wide set of spin glass examples. In this plot, which is presented as a function of the reduced variable $T/T_g\ Ln(t_w/\tau_0)$, the results from 3 samples (from [joh99], see caption of Fig.7) at 2 different temperatures do all fall on the same straight line, which corresponds to a power law behaviour $N_S \sim (t_w/\tau_0)^{0.45T/Tg}$. The number of correlated spins is, as expected, an increasing function of $t_w$, and the numbers reached in the experimental times are ~$10^4 - 10^6$. In the 3 considered samples the spin anisotropy remains rather low, they are examples of "Heisenberg-like" spin glasses. Data from a good example of an Ising-like spin glass [ito86] is also presented in Fig.7 [bert04]; the results lie below the other data by about a factor 10, and do not obey the same simple power law as the Heisenberg-like spin glasses. We have developed elsewhere a comprehensive description of the results from all investigated samples in a common framework [bouchaud01,bert04], which we do not detail here.

Numerical simulations of spin glasses have been conducted intensively by several groups during the last decade [kisker96,komori99,marinari00,berthier02]. In simulations, 4-point correlation functions can be determined, giving a direct access to the dynamical correlation length $\xi$, whereas the experimental procedure yields $N_S$. In the simplest approximation, as was used for the discussion of the experiments in [joh99,bert04], $N_S \sim \xi^3$. However, in more details, the simulations indicate that the equilibrium correlation function involves a power law prefactor $1/r^\alpha$, suggesting that the spins which flip coherently constitute a "backbone" of fractal dimension $d-\alpha$. The calculated values of $\alpha$ are ~0.5 for Ising spins and ~1 for Heisenberg spins, hence yielding $N_S \sim \xi^{2-2.5}$.

Hence, whatever the exact form of $N_s(\xi)$, the $N_S$ values obtained in experiments correspond to values of the correlation length $\xi$ of the order of a few 10 to a few 100 lattice units. This is not very much: due to frustration, spin-glass order grows slowly. Yet, this is still much more than in numerical simulations, where, in units of the elementary time $\tau_0$ which is one Monte-Carlo step for numerics, times were mostly limited to ~$10^6$ (as opposed to $t_w/\tau_0 \sim 10^{13-17}$ in experiments), yielding $\xi$ of the order of a few lattice units only. Very recent progresses have allowed simulations up to ~$10^{10}$ Monte-Carlo steps, yielding $\xi$~10 [belletti08].

Most astonishingly, the general trend of the results from *Ising* spins simulations is a power law behaviour $\xi \sim (t_w/\tau_0)^{0.15T/Tg}$, that is, almost the same power law as in experimental *Heisenberg-like* spin glasses (same exponent if $N_S \sim \xi^3$), although in a very different time range.

On the other hand, there are now a few simulations on Heisenberg spin glasses [berthier02]. They yield a higher $\xi$ than in the Ising case (same trend as in experiments). Hence, the comparison between *real* and *numerical* spin glasses, observed in different time regimes, remains a very puzzling problem. The study of "superspin" glasses (section 3.b below) may offer an opportunity to bridge the gap between the different time regimes.

## 3.b Superspin glasses

Magnetic nanoparticles are single domain objects if small enough (typically, <10nm). For common ferromagnetic materials (Fe, Co, Ni and their oxides) the Curie point is far above room temperature, and the particles may be considered bearing a fixed amplitude giant moment called a "superspin" (~$10^4$ Bohr magnetons per particle). At high temperature the superspins form a superparamagnet, but at lower temperature the flipping of the superspins becomes frozen due to the individual anisotropy energy of the particles (either related to magnetocrystalline anisotropy or shape and surface effects) [dormann97]. If the particles are sufficiently concentrated (typically, more than a few percents volume fraction of magnetic material), due to the strong magnetic moments, dipole-dipole interactions may cause a collective freezing at a higher temperature than that of the blocking of the individual particles [dormann97]. The transition to this frozen state has been characterized as a spin glass-like transition with magnetic non-linearities and critical dynamic scaling. Aging and memory effects can be observed in this superspin glass state (see e.g. [jonsson95,mamiya99,zhang07], and other references in [parker08]).

With the aim of extracting the dynamical correlation length and its age dependence in the superspin glass phase, we have applied the experimental procedure described above (section 3.a) to a frozen suspension of $\gamma$-$Fe_2O_3$ nanoparticles of typical size 8.5nm [parker08] and volume fraction 15% [wandersman08]. We find that the number of correlated superspins grows with a similar power law as observed in experiments on Heisenberg spin glasses and simulations of Ising spin glasses [wandersman08].

At the temperature at which the relaxations were measured (70% of the superspin glass transition at ~80K), the relaxation time for the flipping of the nanoparticle magnetic moment is of the order of 100 microseconds. Hence, in units of this elementary time $\tau_0$, the time range explored in the experiments is $t_w/\tau_0 \sim 10^{5-9}$. This time regime thus brings very interesting informations between that of the numerical simulations (<$10^6$, and even up to $10^{10}$) and that of the experiments ($10^{13}$ to $10^{17}$). The case of [wandersman08] was that of randomly oriented nanoparticles. New experiments are in progress on *oriented* nanoparticles [nakamae09], for which the existence of a unique common anisotropy axis may mimic a situation close to that of the Ising spin glass. We shall then have a possibility of comparison between Ising-like and Heisenberg-like spin glasses in the short (simulations), intermediate (real superspins and some new simulations [belletti08]), and long (real spin glasses) time regimes.

## 4. Conclusion

Since the discovery by G. Parisi of his famous solution to the mean-field approximation of the spin-glass problem, there was a great deal of speculation as to its possible relevance to real spin glasses. The comparison of simple quantities like the FC and ZFC susceptibilities with those obtained from the mean-field theory (MFT) is striking [parisi06]. Below $T_g$, the FC susceptibility of the spin glass samples is essentially flat, as is found in MFT. In contrast, in a perhaps naïve interpretation of the droplet model, one would expect a FC curve which increases towards low temperatures (or tends to do so by some slow relaxation), as is observed in frozen superparamagnets (following a suggestion by N. Sourlas). This is not the case: the flatness of the FC curve below $T_g$ seems to be the hallmark of the spin-glass state, even in the case of magnetic nanoparticles: when comparing ferrofluids of various concentrations, the flatness of the FC curve is found when the interactions are strong enough (high enough concentration) to produce a superspin-glass state [dormann97,jonsson95,parker08].

More sophisticated quantities give other clues towards some relevance of the mean-field description, in terms of the Parisi solution, to the experimental situation. The Fluctuation-Dissipation ratio, measured by D. Hérisson and M. Ocio [herisson02], is clearly in favour of a "non-trivial P(q)", that is, a behaviour similar to the results of MFT with replica symmetry breaking (the question of whether this is rather 1-step or full RSB remaining open for now) [mezard87,franz94,parisi06]. The numerous experiments on the influence on aging phenomena of temperature changes [refregier87,vincent07], namely "rejuvenation and memory" effects, are finely described in terms of a hierarchical organization of the metastable states as a function of temperature, even quantitatively (see in particular [sasaki02]). This hierarchical picture is obviously reminiscent of the hierarchical organization of the pure states as a function of the overlap in the Parisi solution of MFT [mezard87]. Moreover, the analysis of the temperature dependence of the barriers developed by Orbach's group and Saclay [hammann92] have shown a possible link between what is seen of the metastable states in the experiments and the pure states themselves, bringing a support to a picture of a continuous critical line below $T_g$, as obtained in some developments of the MFT [mezard84,mezard87].

The hierarchy of metastable states can now be transcribed in terms of a "hierarchy of embedded correlation lengths" [bouchaud01]. These embedded length scales are measured, in the complementary scales offered by spin glasses [bert04] and superspin glasses [wandersman08,nakamae09]. Work is in progress which will allow developing interesting comparisons with numerical simulations. The obtained length scales do not vary abruptly with temperature, but they allow the occurrence of *independent* aging evolutions at *different temperatures* thanks to a rapid divergence of their characteristic times with temperature [bouchaud01]. Whether these embedded correlation lengths can be understood in terms of droplets is still to be discussed, but in any case we should now be a bit far from the original picture of "compact and independent droplets". Modern views of the droplet model (like [yoshino01,jonsson04]) are certainly able to capture a good part of the experimental features. The question of the relevance of a possible "temperature-chaos" to rejuvenation and memory effects has been amply discussed (see in [bouchaud01,yoshino01,berthier02,berthier03, jonsson04]). One may today consider that temperature-chaos is not a necessary ingredient to a full description of rejuvenation effects, as has been discussed above [sasaki02]. But both types of effects (classical thermal variation of statistical weights and temperature-chaos) may be expected to co-exist. Simulations of the 4d Ising spin glass [berthier02, berthier03] show the interesting example of a system exhibiting strong rejuvenation effects in the absence of chaos, which is not expected to occur below much larger dynamic length scales.

Other issues from the Parisi MFT concern the existence of large-scale excitations of small (non-extensive) energy, that is, in the language of droplets, a vanishingly small value for the exponent $\theta$ which defines the dependence on the size $L$ of an excitation $F$, $F \sim L^{\theta}$. Such low-energy large-scale excitations have been identified in simulations [krzakala00,lamarcq02], with a geometry described as "spongelike". On the basis of the experiments on disordered ferromagnets [dupuis02], we concluded that aging dynamics in the spin glass resembles in many aspects that of pinned walls. The spongelike excitations might be plausible candidates for these mysterious walls, providing us with still another (even if tenuous) similarity between the Parisi mean-field spin glass and real systems.

Let us end this paper with a question which remains vertiginously open: is there an infinity of pure states in the real spin glass, as obtained in the case of the Parisi mean-field spin glass, and in contrast with the 2-state nature of the droplet spin glass ? Does it make any difference in any real situation ? Experiments and numerical simulations seem to be intrinsically limited to exploring the inner of a single pure state, or maybe averaging over several of them. New prospects might come out from experiments on mesoscopic (non self-averaging) samples, as recently proposed in [carpentier08].

*We are grateful to Jean-Philippe Bouchaud for important discussions and a friendly re-reading of this manuscript.*

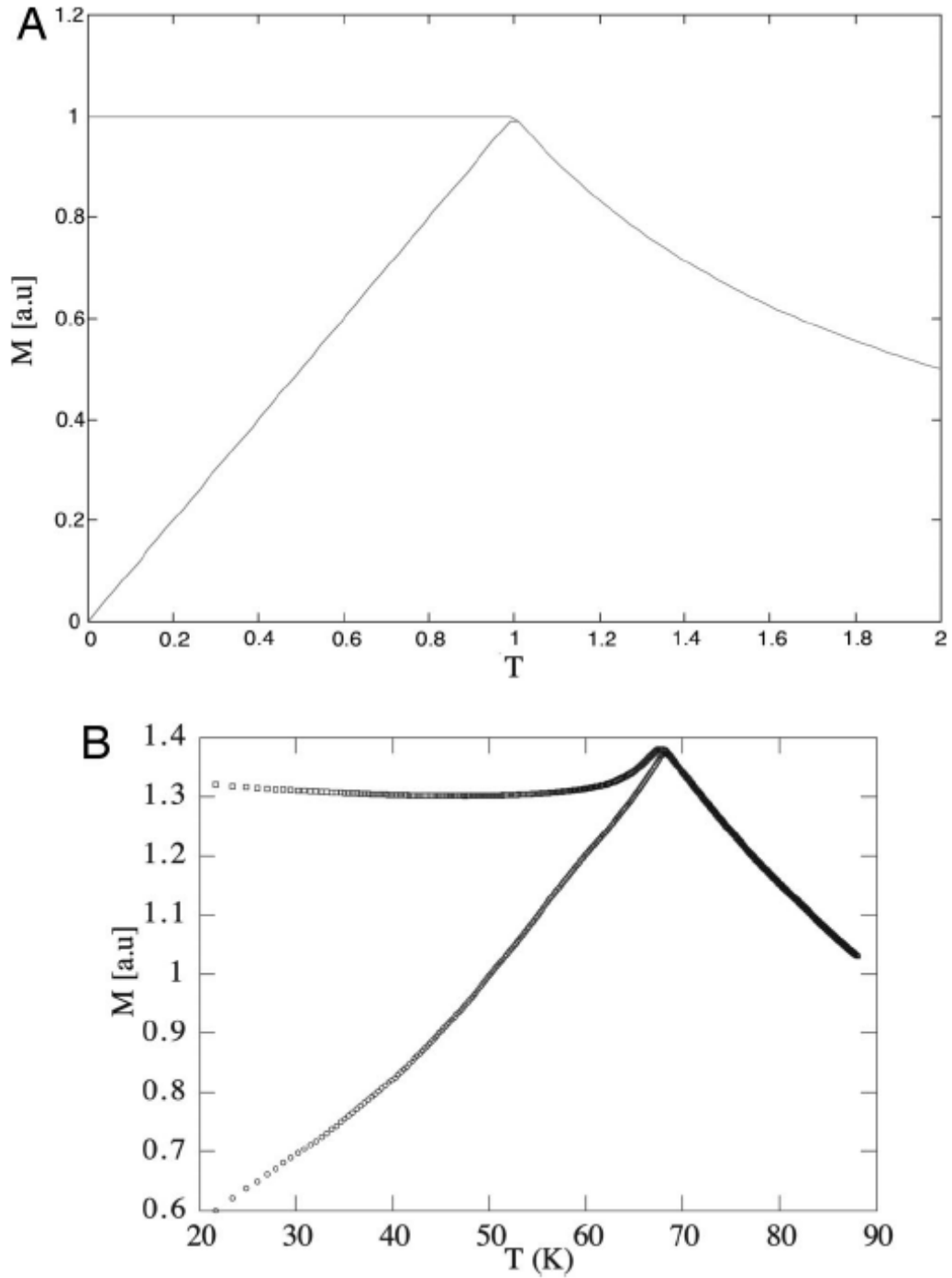

Figure 1: (from [parisi06]) A) Magnetic susceptibility of the mean-field spin glass, obtained analytically [mezard87]: linear response susceptibility $\chi_{LR}$, i.e. response within a state, and equilibrium susceptibility $\chi_{eq}$, thermally averaged over the whole phase space ($\chi_{eq} > \chi_{LR}$). B) Measured FC- and ZFC-magnetisation vs. temperature of the Cu:Mn$_{13.5\%}$ spin glass ($M_{FC} > M_{ZFC}$) [Djurberg99].

Figure 2: "Fluctuation-dissipation plot" (FD plot), combining the results from response function (vertical scale) and spontaneous fluctuation (horizontal scale) measurements by D. Hérisson and M. Ocio on the $CdCr_{1.7}In_{0.3}S_4$ spin glass, at 3 temperatures $T = 0.6, 0.8$ and $0.9T_g$ [herisson02]. The vertical scales are the normalized susceptibility (left) and relaxation function (right). The horizontal scale is the normalized autocorrelation of the magnetization. The full symbols are raw results, and the full line curves above the symbols are the long-time extrapolations of these results.

For each of the 3 sets of results, an indicative value of the mean "effective temperature" $T_{eff}$ (inverse of the mean slope $1/T_{eff}$ of a crude linear fit to the extrapolated data) is given. The straight lines have slope $1/T$, $T$ being the measurement temperature. Hence, each of the 3 sets of results shows the crossover from equilibrium dynamics (slope $1/T$, normal FD relation obeyed) to non-equilibrium dynamics (mean slope $1/T_{eff}$, the FD relation indicates a temperature $T_{eff}$ which is higher than the measurement temperature $T$).

The dashed line is a rough adjustment to the extrapolated results of $\chi(C)=(1-C)^B$ for $B=0.47$ ($B=0.5$ corresponds to the Sherrington-Kirkpatrick model at small $C$ [marinari98]).

For a full description of these experiments see the original papers by Hérisson and Ocio [herisson02].

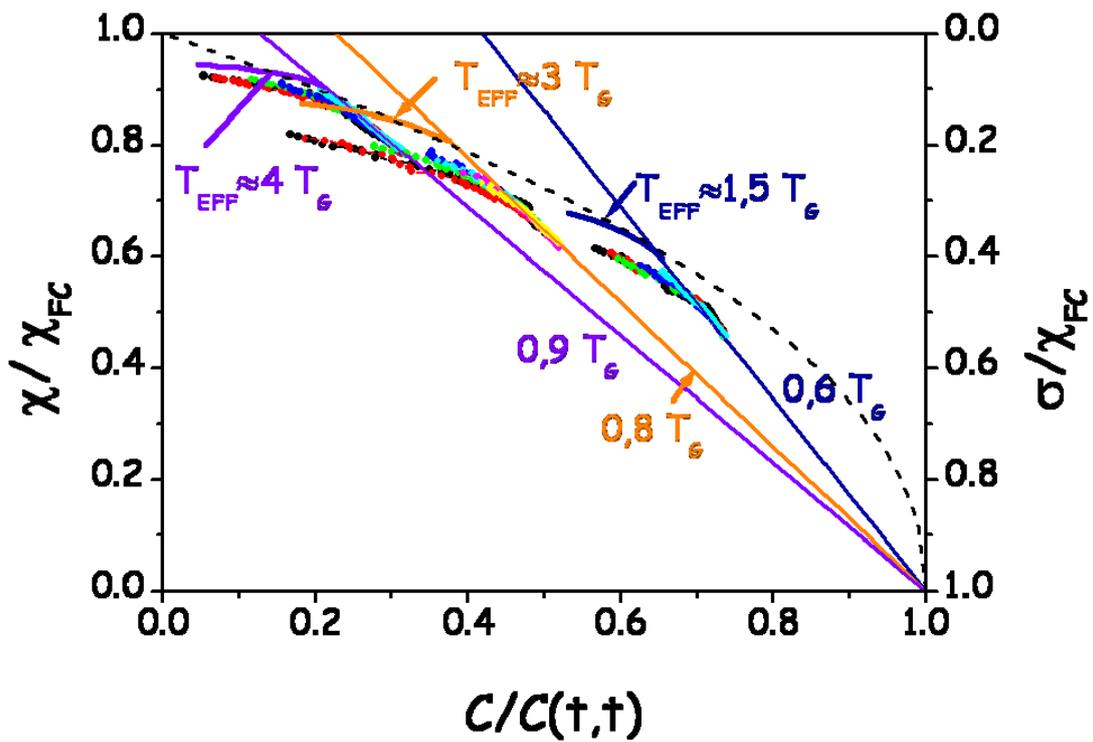

Figure 3 : (from [parisi06]) Left-hand side: probability distribution *P(q)* of the overlap *q* between two equilibrium configurations, in three different cases A, B and C. Vertical arrows correspond to delta functions. Right-hand side: corresponding relaxation function *S* versus autocorrelation *C*, in the same three cases.
Case A: "Classical" hysteresis (e.g. ferromagnet).
Case B: Spin-glass models with 1-step RSB. This case may correspond to structural glasses.
Case C: Mean-field spin glass models with full RSB (e.g. the Sherrington-Kirkpatrick model).
See details and references in [parisi06].

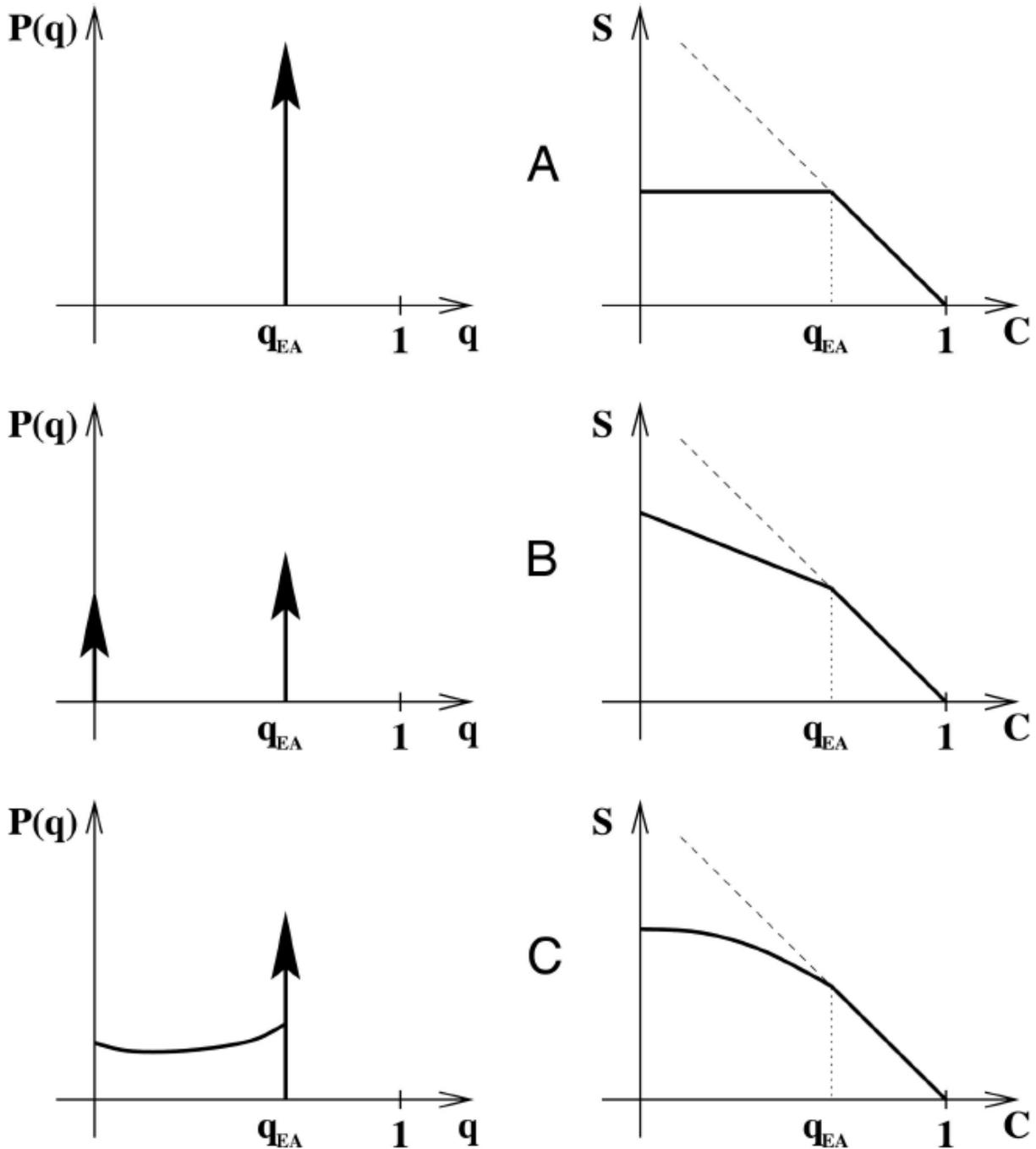

Figure 4: **A)** Example of multiple rejuvenation and memory steps, seen in a measurement of the out-of-phase component $\chi''$ of the ac susceptibility in the CdCr$_{1.7}$In$_{0.3}$S$_4$ spin glass sample [bouchaud01]. The inset sketches the temperature variation procedure during the experiment. The temperature $T$ was decreased by *2K* steps, with an aging time of *2000s* at each step (open diamonds). Continuous reheating at *0.001K/s* (full circles) shows memory dips at each temperature of aging. **B)** Schematics of the hierarchical structure of the metastable states as a function of temperature [dotsenko85,refregier87,vincent07].

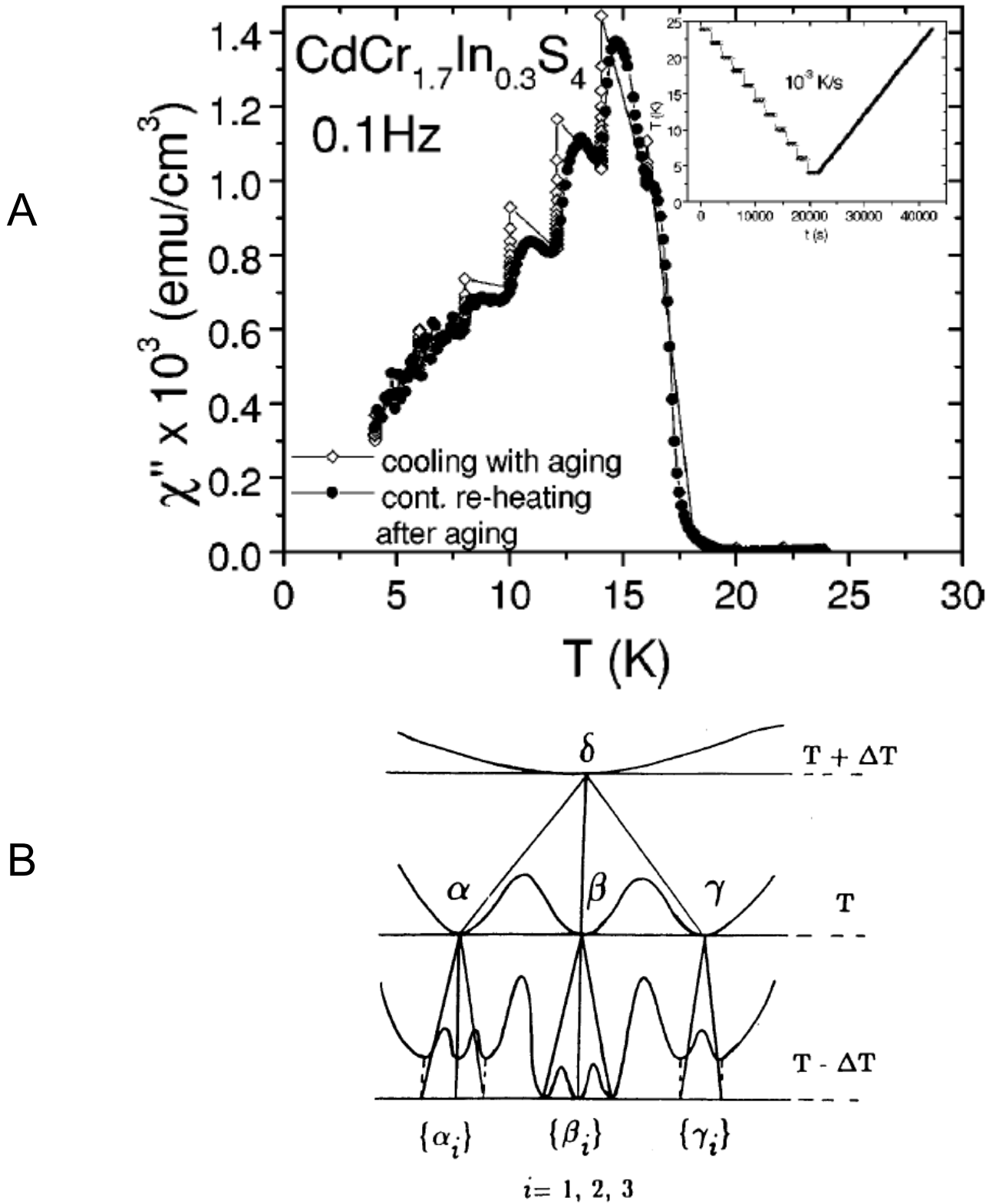

Figure 5 : Rate of thermal variation of free-energy barriers as a function of their height *Delta*, as deduced from the effect of temperature variations on the TRM relaxation of an Ag:Mn$_{2.6\%}$ spin-glass sample [hammann92]. The rate is actually negative, the vertical scale is its absolute value. The barrier heights are normalized to $T_g$=10.4K. The dashed line is a fit to *d(Delta)/dT$_r$=a.Delta$^n$*, with *a=2.9 10$^7$* and *n=6*.

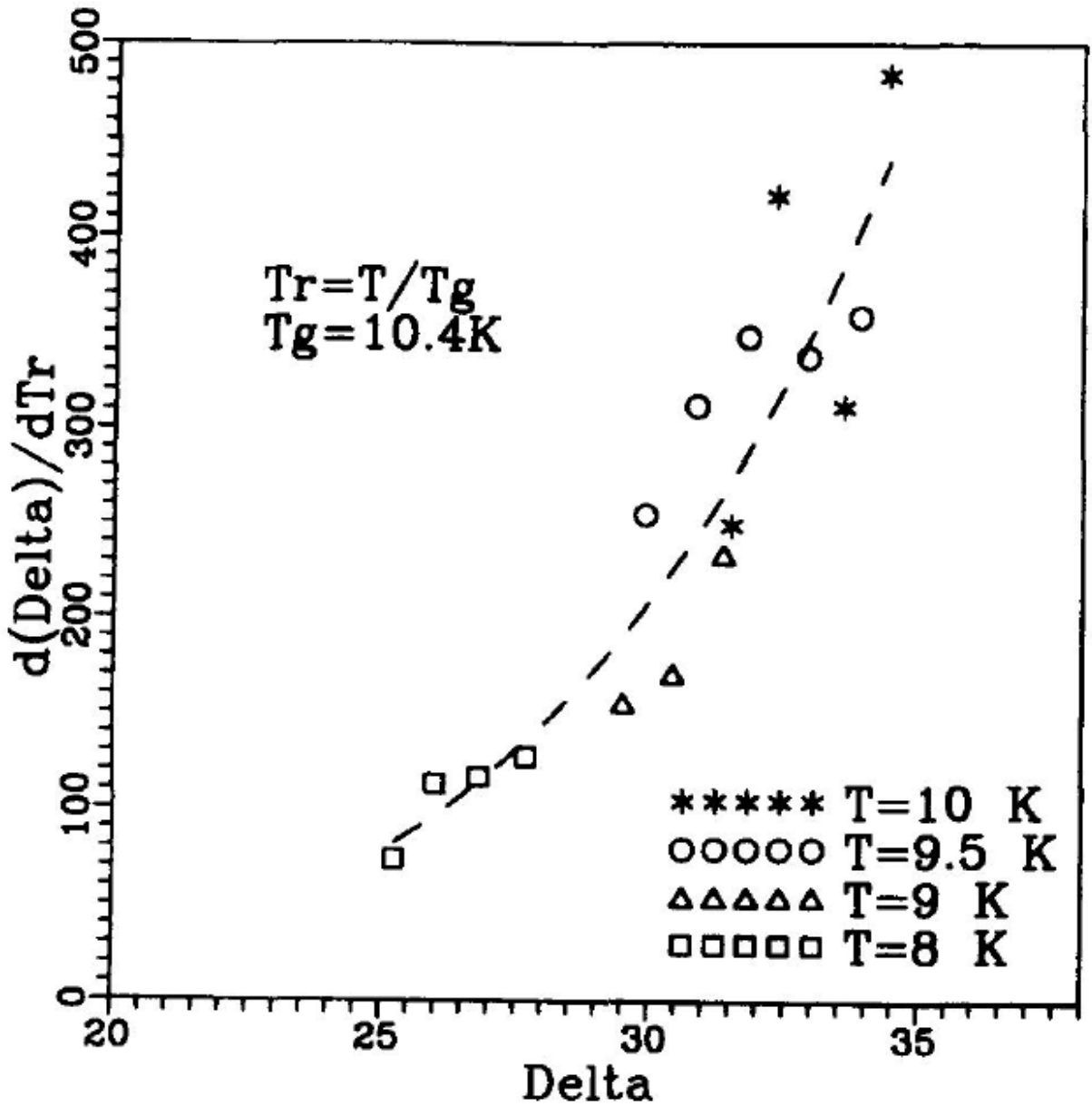

Figure 6 : Sketch of rejuvenation and memory phenomena in terms of the dynamics of an elastic line in pinning disorder. At fixed temperature $T$ after a given waiting time, the line matches the pinning site over a typical distance $L_T^*$. Smaller length scales are fluctuating (fuzzy drawing in the sketch). As the temperature is lowered to $T-\Delta T$, the configurations corresponding to small scale $L_{T-\Delta T}^*$ details ($L_{T-\Delta T}^* < L_T^*$) are no more equivalent, the line has to find the most favourable configurations among small scale pinning disorder, giving rise to new dissipative processes (rejuvenation). Rejuvenation occurs while the memory of reconformations at larger length scales $L_T^*$ is preserved (from [bouchaud00]).

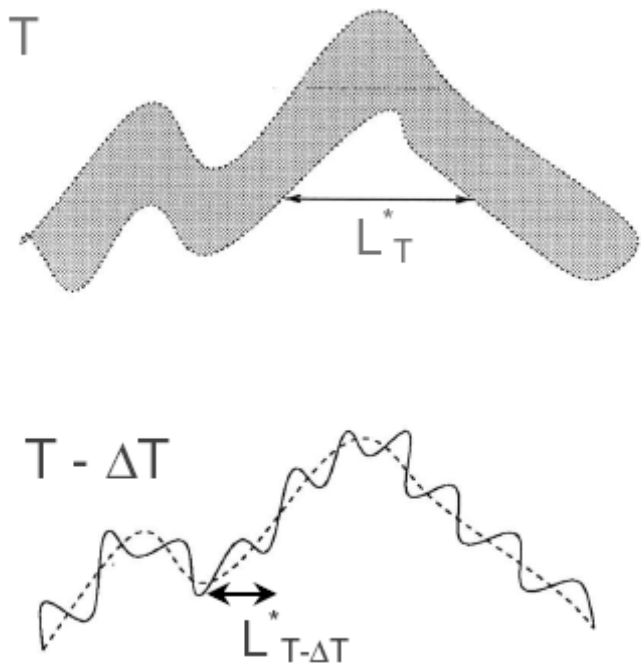

Figure 7 : Number of correlated spins extracted from field variation experiments, as a function of the reduced variable $T/T_g \ln(t_w/\tau_0)$. The points with error bars correspond to Heisenberg-like spin glasses [joh99], they are well fitted by the straight line $N_S \sim (t_w/\tau_0)^{0.45T/Tg}$. The full circles lying below the other points are from the $Fe_{0.5}Mn_{0.5}TiO_3$ Ising sample [bert04].

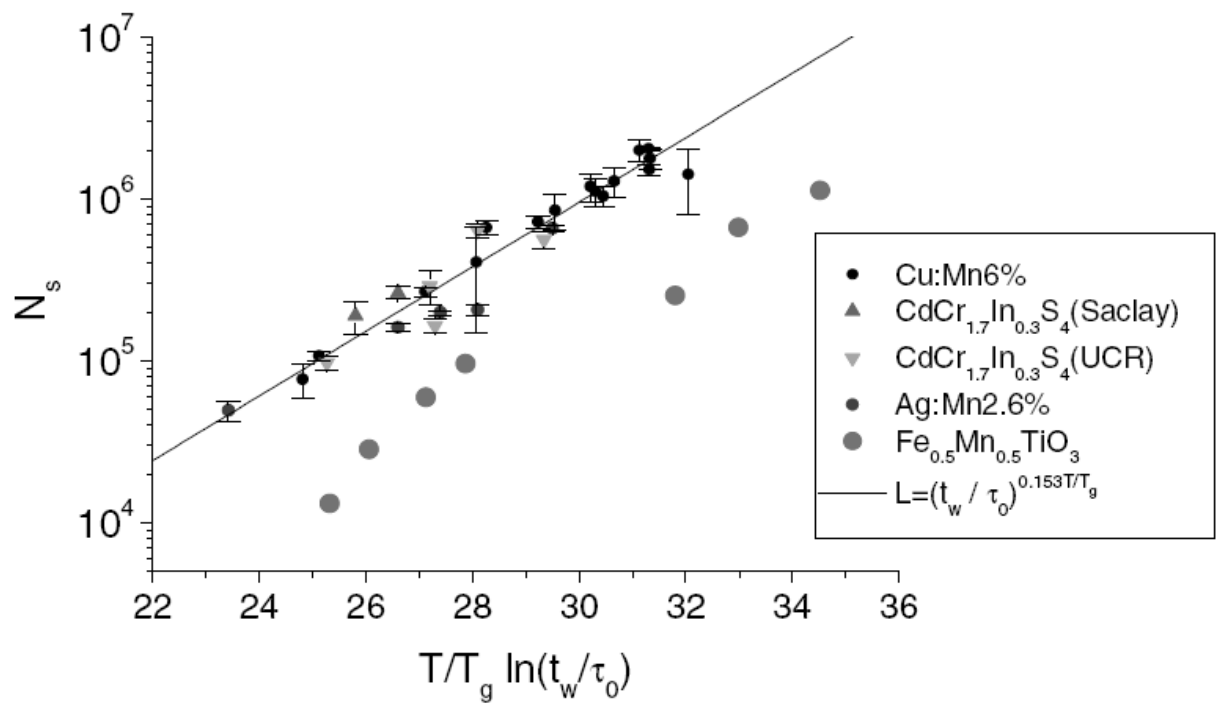